# *The Applications of Subsurface Assemblies on Thermoelectric Generators for Lunar Processes*

Eliazar Montemayor III
*Department of Engineering, Physical, and Computer Science*
*Montgomery College*
Rockville, USA
ORCID: 0000-0001-8241-2444

Matthew Lewton
*College of Engineering*
*Purdue University*
West Lafayette, USA
ORCID: 0000-0001-8339-1198

Kane Timlin
*Department of Engineering, Physical, and Computer Science*
*Montgomery College*
Rockville, USA
kanetimlin@gmail.com

Teresa Doley
*Department of Engineering, Physical, and Computer Science*
*Montgomery College*
Rockville, USA
teadoley@gmail.com

***Abstract*—To take advantage of the distinctive nature of the lunar environment with respect to heat flow due to the low thermal conductivity of lunar soil and the moon's negligible atmosphere, thermoelectric generators have been identified as a potential power source to support self-sustaining processes on the lunar surface. These power-generating systems will take the form of top-down heat transfer systems, relying on soil as the heat sink to maintain significant temperature differentials. Installation of increasingly complex subsurface assemblies to redirect subsurface heat flow have demonstrated strong capability in both enhancing power output and limiting near-surface temperatures.**

## I. Introduction

The future of scientific achievement lies in space exploration and colonization. With an international collaboration between space exploration agencies and companies, the next steps in space exploration begin with remote and later human-manned lunar expeditions to gather data on the feasibility of life on the moon, gradually building the foundations of a habitable lunar colony and establishing a permanent human presence on the moon [1, 2]. However, as these operations become increasingly long-term, there follows a growing need to maintain support resources, such as the development of power systems on the moon.

One such proposed source of power comes through heat. Utilizing the principles of the Seebeck effect, the thermoelectric generator (TEG) takes advantage of an applied temperature differential between the "hot side" and the "cold side" to generate electrical current.

These solid-state devices can potentially be used in conjunction with lunar soil to produce usable energy. Due to the moon's negligible atmosphere and the thermally insulative behavior of lunar soil, there exist natural temperature differentials between the lunar surface and subsurface that can be used to power TEGs.

Exploiting this differential is more complicated than simply resting the TEG atop the surface. Because of this, subsurface assemblies that can force stronger-intensity heat flux and stabilize with high temperature differentials between the TEG's top side, acting as the "hot side," and bottom side, acting as the "cold side," must be installed to efficiently convert heat to usable electricity. Instead of having the lower-conductivity lunar surface act as the heat sink, the way a TEG placed directly on the soil would, subsurface assemblies effectively lower the cold side of the TEG to lower depths below the surface to place it in contact with higher-conductivity and lower-temperature soil.

This paper investigates the ways in which different types of assemblies underpinning the workings of thermoelectric technology under such an environment can affect heat flow. While numerous studies have already detailed the generalized workings, structure, optimization, and design of TEGs, this paper aims to put this technology into practice. The equipment developed from this work allows for the dual purpose of recycling otherwise-wasted heat back into the operating cycle of in-situ equipment as well as providing a natural passive heat exchanger by which heat can be dissipated into the environment, away from any operating surface-level devices.

## II. Background

On Earth, atmospheric convection greatly affects near-surface temperatures: any form of heating due to solar radiation is significantly reduced by the atmosphere, and the surface is cooled by atmospheric convection. However, the moon operates in a near-vacuum state with minimal atmospheric densities. Because of this, most thermal energy on the moon's surface must be dissipated through conduction into the soil.

Lunar soil is characterized as a "silty sand" because of its fine grain size, low density, and elastic behavior. It has a thermal conductivity of approximately 0.003−0.009 W/mK [2, 3, 4].

The Apollo 15 and 17 drilling experiments found that temperatures 35 cm or lower beneath the lunar surface remain

This work was supported in part by the National Aeronautics and Space Administration through the 2021 NASA MINDS program under Cooperative Agreement 80NSSC20M0194 and in part by Montgomery College under authorization by the Dean of Science, Engineering, and Technology.
*(Corresponding author: E. Montemayor.)*

E. Montemayor is also with Walter Johnson High School, Bethesda, USA (email: eqmontemayoriii@gmail.com; emontema@montgomerycollege.edu). M. Lewton is also with the Department of Engineering, Physical, and Computer Science, Montgomery College (email: matt.lewton9@gmail.com; mlewton@purdue.edu).

stable throughout the day and are not subject to the monthly temperature fluctuations found on the moon's surface [5]; prior research and analysis suggests that the thermal difference between the surface and subsurface at the equator could be up to 150 K, reaching a stable temperature of 240 K at depths beneath 20 cm [3]. Surface temperatures at the equator are more extreme and depend on the sun's influence, reaching a high of over 400 K and dropping to a low of 100 K. While temperatures at the lunar south pole may not maintain the same values or temperature swings, a similar relationship between its surface and subsurface is expected to persist.

In general, there is an inverse relationship between thermal conductivity and temperature differential; this means that a material with a higher conductivity will result in a smaller temperature differential across its two ends when heat is transferred across it.

The relationship between lunar soil thermal conductivity and temperature for both the dust and soil layers on the moon can be modelled mathematically [3], as it has been found that thermal conductivity increases with bulk density, which itself increases with soil depth.

Each instantaneous increase in soil depth results in an increased uniform bulk density within each successive layer and, correspondingly, increased thermal conductivity. As conductivity of each successive layer of soil increases, the differential must decrease in turn. As the density and conductivity increase, the decreasing differentials cause the temperatures to stabilize at a depth of 20 cm.

## III. DESIGN

### A. General Assembly

In this scenario, the application of a TEG forms a top-down system of power generation by utilizing a heat source above and lunar soil as a heat sink below. By establishing the heat sink below, where the bottom side of the TEG is directly or indirectly in contact with lunar soil, the TEG can rely on the natural conduction of the soil with the moon's essentially limitless heat capacity to maintain the temperature differential. However, if the TEG were to be placed atop the soil without such a modification, the maximum differential it could achieve would depend on the layer of soil coplanar to its bottom face. Thus, the use of a subsurface assembly is required to achieve greater effects.

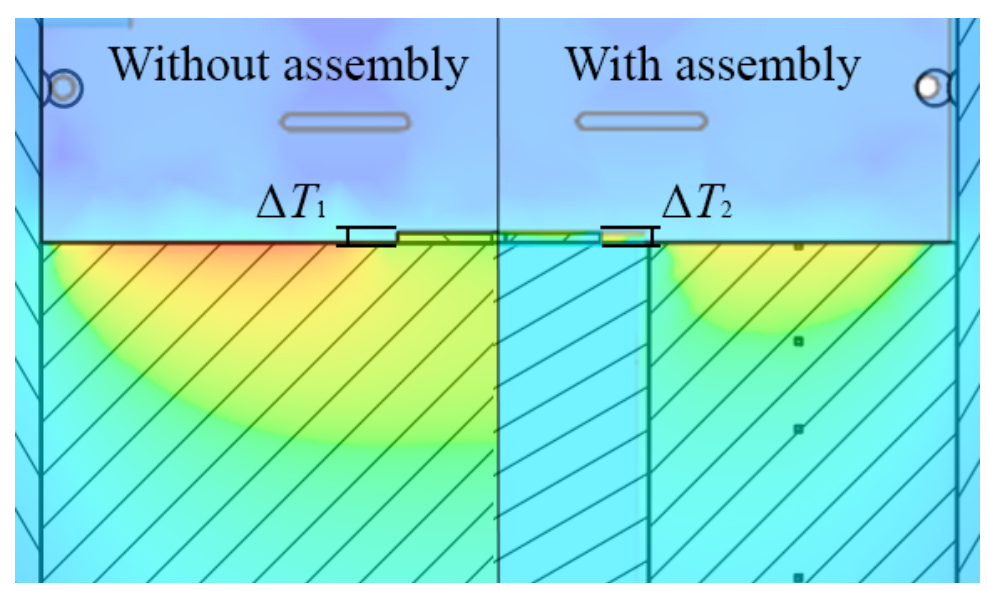

Fig. 1. Heat map of a general design showing heat distributions with the TEG resting on soil directly (left) and of the TEG resting on an assembly (right).

The defining feature of an assembly is its capacity to limit horizontal heat flow and to allow the TEG to utilize lower depths with higher thermal conductivities and lower temperatures. Consider the case of a hypothetical, cylindrically symmetrical assembly composed of a thermal conductor with infinite conductivity and a thermal insulator with zero conductivity, as pictured in Fig. 1. Were the TEG placed as it is without the assembly, its maximum temperature differential would be $\Delta T_1$. The thermal conductor effectively bridges the bottom side of the TEG to a regolith layer at the bottom of the conductor, allowing for optimized heat flow and therefore maximized temperature differential $\Delta T_2$ induced by the heat sink, a differential much higher than the initial value. Heat flow at such a distance is facilitated by the thermal insulator which completely negates horizontal heat flow from the surrounding sand into the conductor.

A conductor made of a homogeneous material but has non-infinite conductivity will develop a small and negligible temperature gradient. In practice, the insulation does not completely block horizontal heat flow as insulative materials still have some amount of conductivity; care should be taken to limit horizontal flow as much as possible.

Increasing the contact surface area between the thermal conductor and low-temperature soil, such as by extending it deeper past the length of the insulation, increases the intensity of the outgoing heat flux in Fig. 1 relative to a system with lower contact area, causing temperature changes relative to the previous system as the incoming and outgoing heat fluxes reconcile to thermally stabilize the system.

### B. Experimental Assembly

To model a possible assembly design using available materials and resources, a proof of concept model was built to test whether the proposed system can effectively produce a temperature differential and electric power in a simulated lunar environment.

The prototype experimental assembly was built as a modified helical anchor—though there are many other possible designs that differ from the one chosen here—as such load-bearing objects have been proposed as ground anchoring systems for lunar civil structures. Helical anchor dimensions were sourced from Led Klosky's *Behavior of Composite Granular Materials and Vibratory Helical Applications*, which recommends specific sizes for these objects in the context of deep-foundation support on lunar regolith [4].

The assembly is structured in four distinct metal parts: a steel pipe for the helical anchor, two concentric copper "radiator cylinders" that sit inside the anchor, and a copper block for the heat transfer block (HTB).

The thermoelectric module rests on the HTB, which facilitates heat transfer between the thermoelectric module above and the radiator cylinders beneath. The radiator cylinders are made of conductive metal and act as radiator fins that sit in a concentric fashion. The steel pipe is insulated along its inner radius at depths above 20 cm to prevent heat in-flow from the sides, only allowing for a strict top-down transfer of heat. In actual use, the HTB can be extended above the surface as far up as necessary to reach an external heat source. The TEG module consists of four 40 mm commercially-available square TEGs that sit on top of the HTB. The remaining exposed surface of the HTB is covered with a layer of insulation.

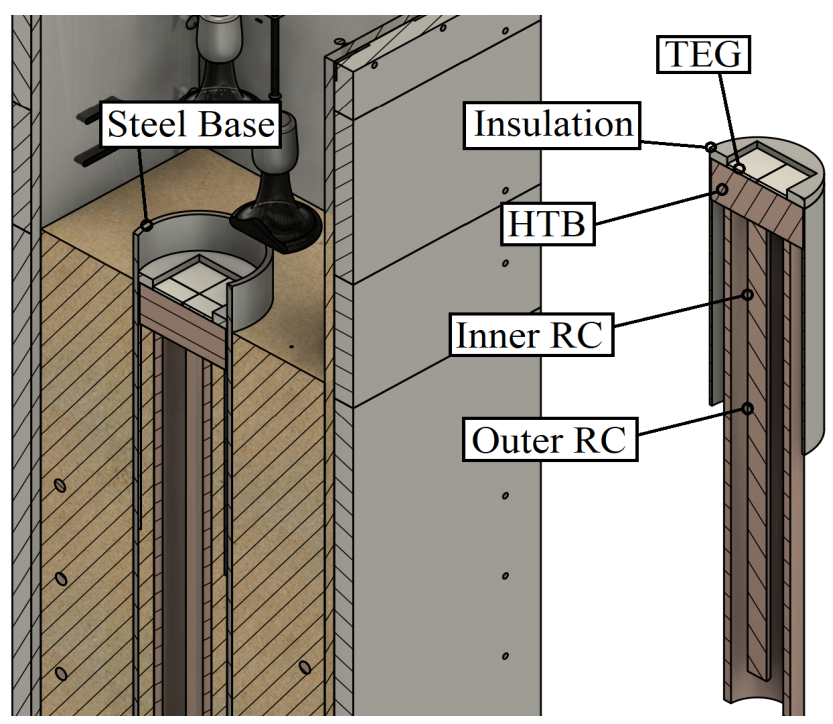

Fig. 2. Cross-sectional image of the experimental assembly.

To thoroughly test the effectiveness of the concept and design, the results of the assembly were investigated with a combined system of analysis using CAD modelling and experimental testing. Utilizing both simulated programs and actual experimentation allowed for a working basis of the design's thermodynamic capabilities in idealized simulations, to be compared with experimental results that account for the irregularities inherent within in-situ operations.

*C. Testing Rig*

The testing rig simulates three key components of the moon's thermodynamic environment: the lunar atmosphere is a vacuum, the surface receives heat as solar radiation, and the soil acts as a heat sink and stabilizes in temperature at 20 cm.

The lunar vacuum is simulated by an acrylic chamber kept at a near-vacuum state to eliminate convective heat transfer at and above the surface. The source of heat is simulated by heat lamps situated above the soil's surface, emitting radiative heat directed at the soil surface. To simulate the heat sink-like nature of lunar soil with its effectively infinite heat capacity, an external liquid cooling system is embedded in the testing rig. The coolant flows through coiled copper pipes below the 20 cm depth. The cooling system maintains room temperature of the coolant, removing however much heat is applied to the system by the heat source.

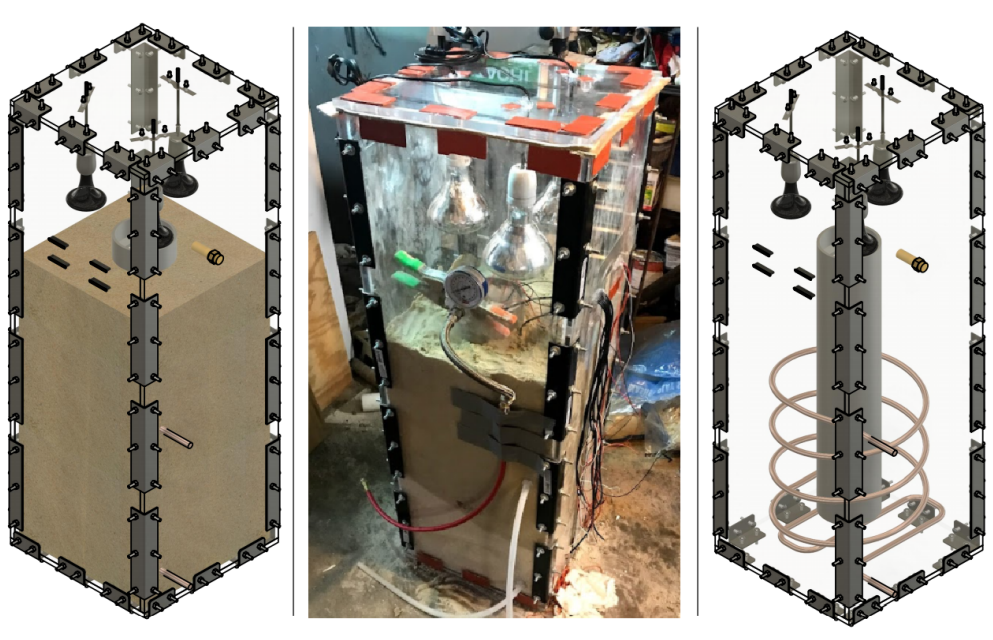

Fig. 3. The experimental testing rig with CAD models and in-situ photos.

The lunar soil is simulated with dry sand, which mimics lunar regolith's small grain size, abrasiveness, and other physical properties. The entire testing rig is insulated with ceramic fiber insulation and the half-inch acrylic sheets that make up the vacuum chamber.

To measure voltage, power from the TEG modules runs in series through a 9.4 kOhm applied resistance. Total internal resistance of the TEGs was measured at 60 Ohms and is considered negligible when in series with the applied resistance, given the low current. The measured voltage drop approximates TEG open circuit voltage.

*D. Experimental Design*

Five sets of experiments and six sets of simulations were conducted with the experimental assembly and testing rig, with the complete assembly being installed into the rig gradually with each successive set. The first set, the Control set, consists of only sand; this is meant to provide a basis for what the temperatures would look like before installation. The full experimental assembly is split into five variants, each of which was tested as its own set. Assembly 1 consists of sand with TEGs resting atop of it. Subsequent sets include increasingly complex assemblies with the addition of the insulated HTB and inner radiator cylinder in Assembly 2, an uninsulated outer radiator cylinder in Assembly 3, full outer insulation in Assembly 4, and the steel pipe in Assembly 5.

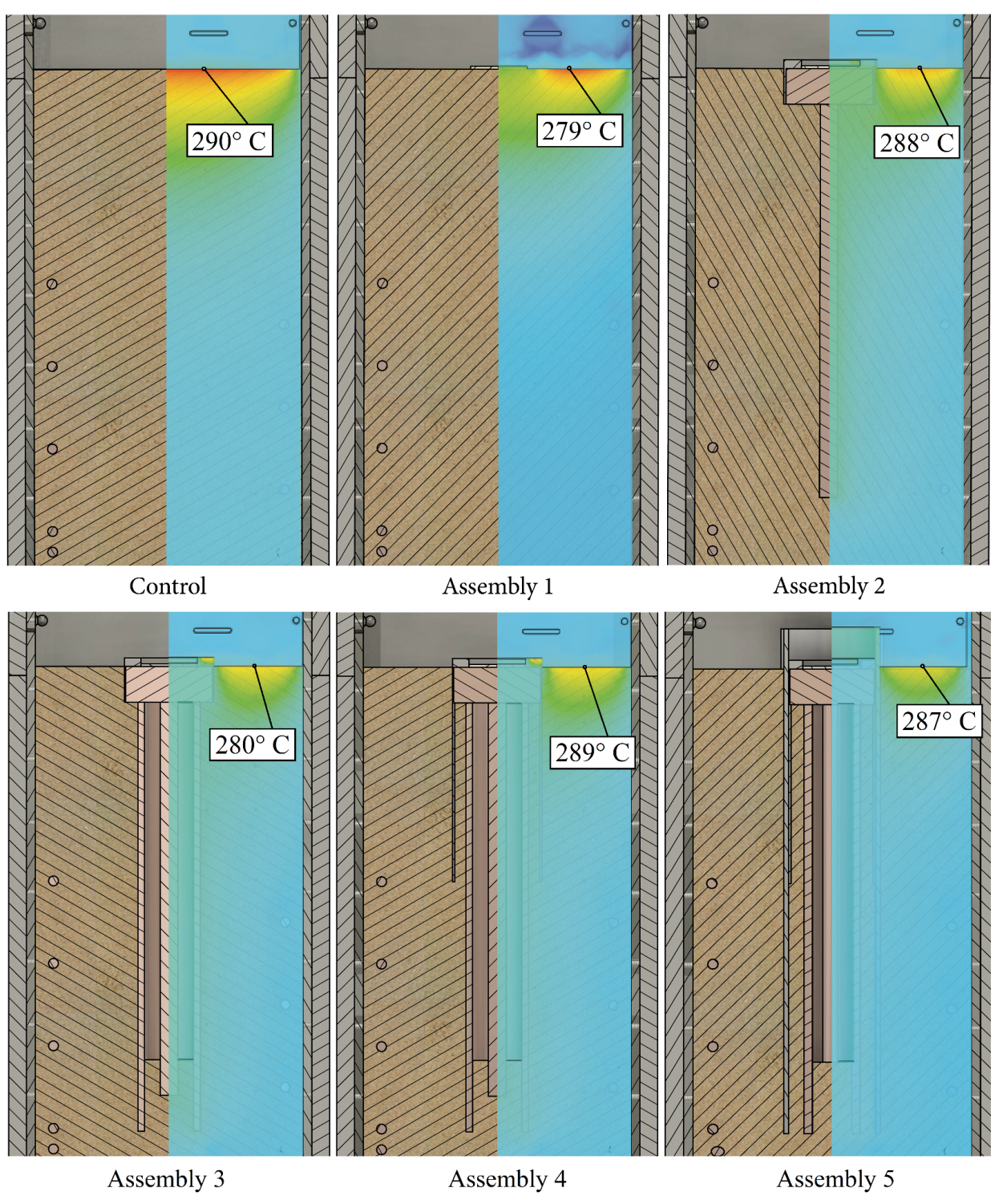

Fig. 4. Cross-sectional images of assembly sets with heat maps of 750 W highlighting simulated maximum surface temperatures.

Thermal simulations were conducted using Fusion 360. The parameters of the simulations were designed to model the conditions of the actual experiments as follows: all power from the heat lamps was applied evenly across the sand surface and exposed portions of the TEG assembly; all heat from the heat lamps was absorbed by these exposed surfaces via radiation with emissivities appropriate for the materials; all materials start at a room temperature of 20° C; the cooling system maintained a temperature of 20° C and could remove up to 1500 W from the system; and the thermal insulation outside the testing rig was cooled by convection with the still ambient air at 20° C. While the experiments were conducted at inner stabilized pressures of 6–20 kPa, their corresponding simulations were run under the assumption of a perfect vacuum state.

While Fusion 360 cannot account for the heat removed as electricity by the Seebeck effect, it does provide an indication of how much heat would transfer across the assembly otherwise.

## IV. RESULTS

### *A. Simulation Results*

The results from simulations and experiments mirrored expected lunar soil temperature values. Temperatures reached their maximum values at the surface of the simulated sand, which then declined sharply even before reaching 20 cm in depth—a similar drop-off to what is expected for true lunar soil. As expected, the thermal gradient of the system stabilized in a top-down manner and supported the predicted direction of heat flux in a true lunar system. The highest temperature values are observed at the top of the system, with numbers decreasing exponentially as depth increases up to their minimum values at the bottom of the testing rig, and the lowest temperatures can be observed at the lowest point of the system; these observations come from scaled heat dissipation of the installed cooling system.

Graphed below are select data points for the hot side and cold side temperatures for simulations of Assemblies 1, 2, 3, 4, and 5.

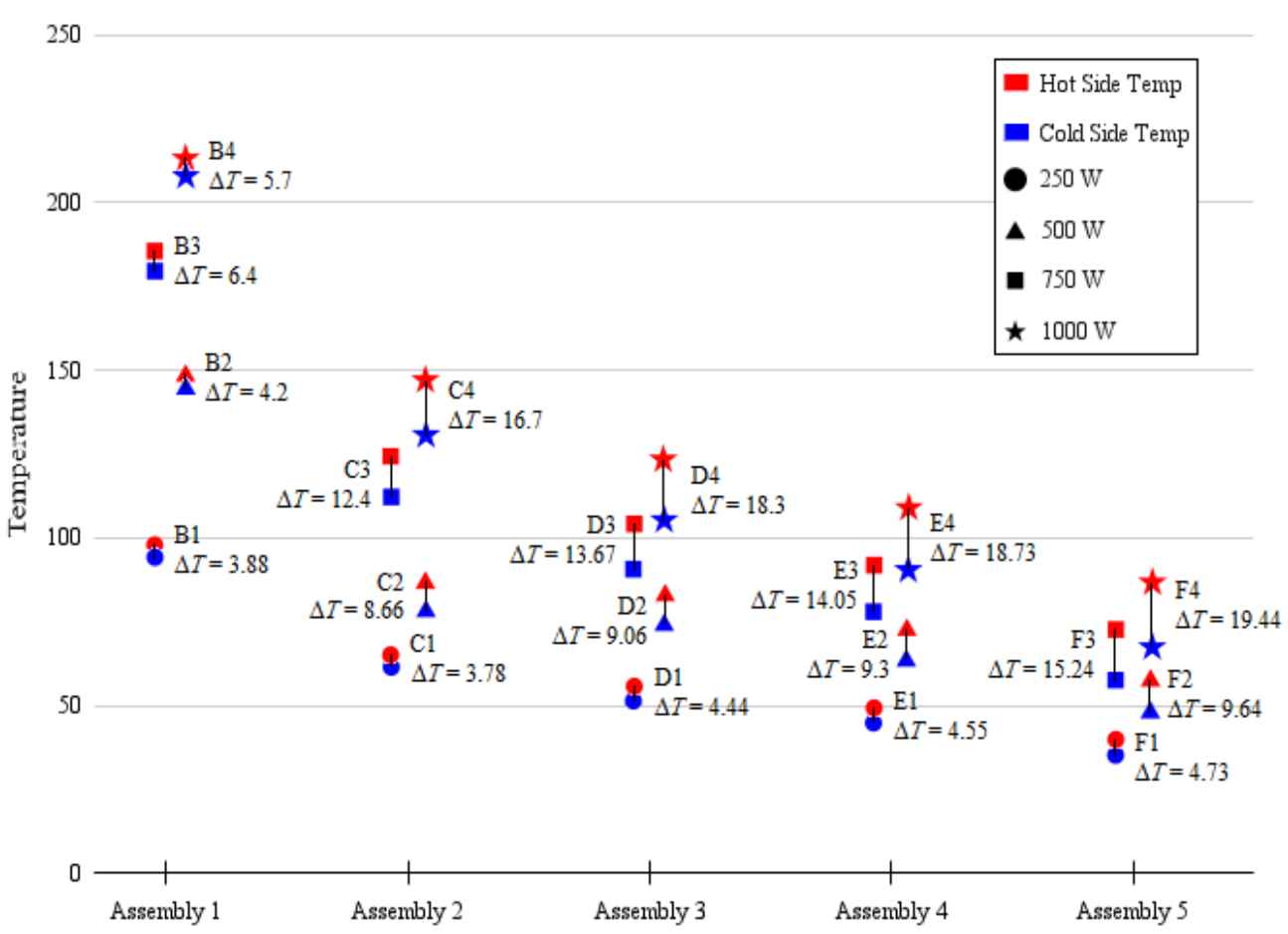

Fig. 5. Stabilized temperature values (in ° C) for the TEG hot and cold plates as measured during simulations.

As expected, increasing "complexity" from each successive assembly caused markedly decreased temperatures and widening temperature differentials. The most effective thermal conductors are defined by their high overall conductivity and high contact surface area with deep soil, which gives rise to high outgoing heat fluxes.

The effectiveness of a subsurface assembly is most strongly demonstrated in the sharp differences between Assemblies 1 and 2, due to lower contact surface area in the former as the latter included an outer radiator cylinder that increased contact area significantly. High temperatures and low differentials are observed in Assembly 1, strongly contrasting the low temperatures and high differentials presented in Assembly 2.

While Assemblies 3 and 4 have the same conductor configuration, the lengthened insulation in Assembly 4 prevents horizontal heat bleed for deeper layers. However, this change is less pronounced than with other changes, indicating that the length of insulation has a diminishing effect in higher depths. The significance of insulation is most pronounced near the surface.

The increase in input heat power within a set affected the resulting temperatures and differentials linearly. The increase in temperature differential across the TEG, with the exception of Assembly 1, indicated that the hot side temperature grew at a faster rate than the corresponding cold side temperature with each increase in input heat.

While heat is less concentrated on the surface with more complex assemblies, especially in the areas immediately surrounding the assemblies, maximum surface temperatures did not seem to be affected, as shown by the heat maps in Fig. 4. While subsurface temperatures near the assemblies were influenced by the corresponding changes in heat flux, heat distributions gradually returned to their expected differentials at larger depths.

### *B. Experimental Results*

During the course of each experiment, temperatures recorded at different locations gradually diverged from room temperature as the input heat caused a top-down temperature gradient across the system. Recorded temperatures over time increased logistically in regions closer to the surface, with decreasing gradients as heat fluxes gradually stabilized, but increased at near-constant rates for deeper layers. Each experiment ran for 1–3 hours before stabilizing.

Some recorded experimental data were omitted from the final results because the commercial TEGs and temperature sensors were rated up to a maximum operating temperature of 125–150° C, and, as such, values that exceeded those limits could not be observed accurately.

Experimental readings validate the relationship between voltage and temperature differential for the TEGs through the Seebeck effect [6]. The growth in voltage mirrors the change in temperature differential during the runtimes of each experiment [7].

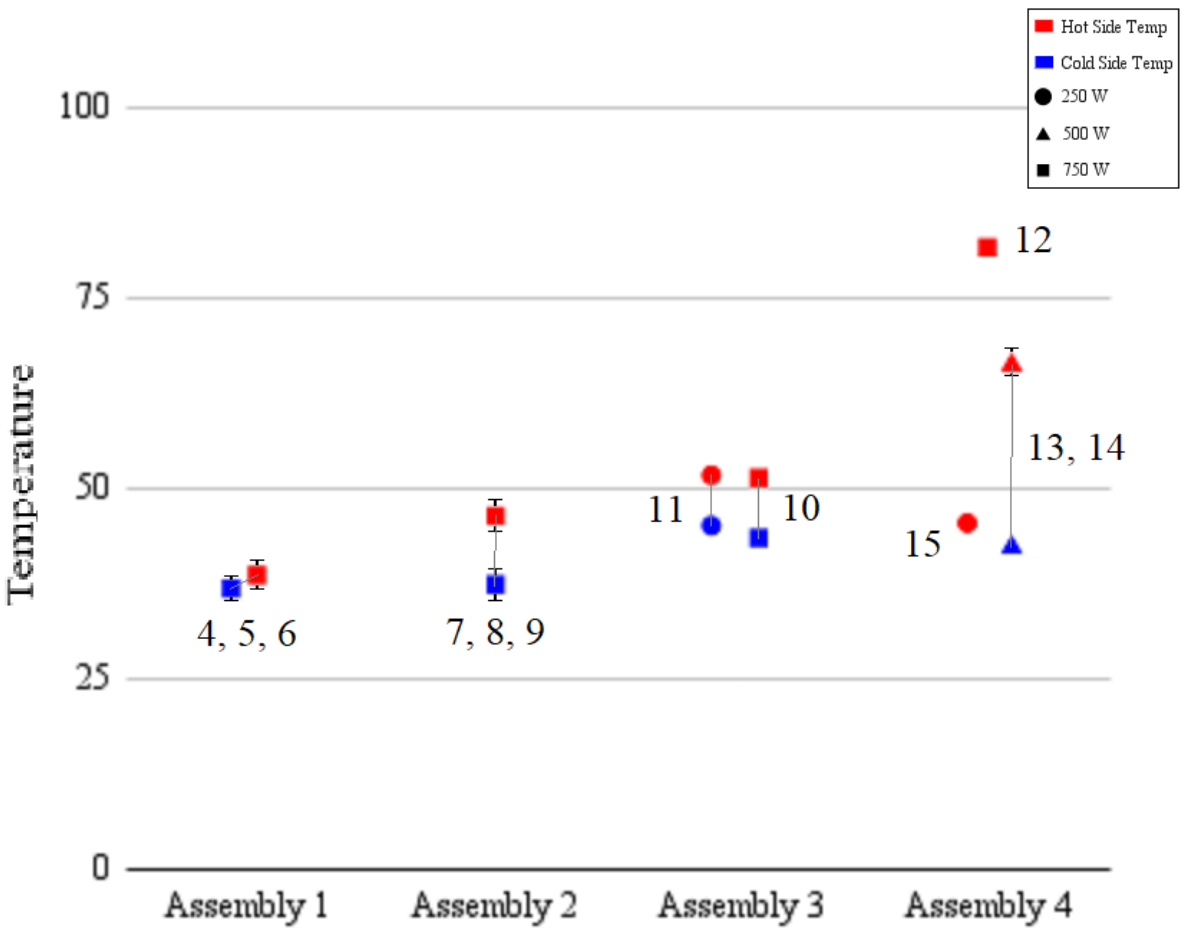

Fig. 6. Stabilized temperature values (in ° C) for the TEG hot and cold plates as measured during experiments.

TABLE I. EXPERIMENTAL RESULTS

| Assembly No. | Exp. No. | Input Heat (W) | Mean Hot Temp. (° C) | Mean Cold Temp. (° C) | Mean Temp. Diff. (° C) | Mean Voltage (V) |
|---|---|---|---|---|---|---|
| 1 | 4, 5, 6 | 750 W | 38.60 | 36.84 | 1.76 | 0 |
| 2 | 7, 8, 9 | 750 W | 46.37 | 37.38 | 8.99 | 1.3 |
| | 10 | 750 W | 51.38 | 43.5 | 7.88 | 0.4 |
| 3 | 11 | 250 W | 52.38 | 45.75 | 6.63 | 0.9 |
| | 12 | 750 W | 81.63 | --* | --* | 0.9 |
| | 13, 14 | 500 W | 66.19 | 42.25 | 23.94 | 0.85 |
| 4 | 15 | 250 W | 45.44 | --* | --* | 0.7 |

*Some cold side values were not recorded due to a technical malfunction.

Experimentally observed temperature differentials and raw hot side/cold side temperatures increased with each successive assembly, as opposed to the simulations' predicted increasing and decreasing trends, respectively. This discrepancy is explained by issues caused by the resetting of the testing rig for each new set of experiments. In the first two assemblies, the sand remained firmly placed and untouched. However, each subsequent assembly required significant sand displacement to allow space for each new insertion, especially with the addition of the outer radiator cylinder and the lengthened insulation in Assemblies 3 and 4. As with lunar soil, the thermal conductivity of sand depends on bulk density, and thus on soil pressure. Significant sand displacement with each reset disrupted this, causing sand to be less compact and therefore less conductive with each set. Issues also arose with the testing rig's inner atmosphere, as the experimental testing rig was not maintained in a perfect vacuum, possibly causing heat to build up in the air.

Owing to the logistical concerns that arose during the resetting process, Assembly 5 was not tested experimentally.

The experimental model had much lower overall temperatures than the simulated model at the same wattages. This is because the experimental model had more effective temperature stabilization due to variable thermal conductivities in the sand. Overall temperature differentials, however, were consistent with the simulated values, suggesting that the internal assemblies functioned similarly to their simulated counterparts.

Mean voltages recorded through each experiment measured mostly within $\pm$50% of expected voltages presented by the manufacturer; given the vague specifications, the inherent inefficiency when testing commercial products at the peak of their intended operating temperature ranges, and the possible effects of changed atmospheric pressure on efficiency, a wide margin of error was expected. Initial experiments with the Control set and earlier assemblies heated the interior of the testing rig to temperatures well beyond the ideal, likely damaging the equipment and facilitating discrepancies in later results.

### *C. Supplementary Testing*

Further simulations using variants of the designs presented in this paper found that the main limiting factor for the resulting temperature differential was the thermal conductivity of the TEG. The area of contact with the TEG is where the HTB is most vulnerable to heat flow as that area warms the entire radiator cylinder system. Increases in the TEG's conductivity were found to correlate with increases in both system temperature and differentials.

They also revealed some inefficiency, caused by the TEG module's square shape, which develops in more complex assemblies. The inconsistent width of the TEGs with respect to the central axis of the system and the perimeter formed by the surrounding thermal insulation leads to uneven heating of the TEGs. This demonstrates that the optimal TEG geometry for this system is circular in shape and centered around the central axis, with a cylindrical wall of insulation.

To observe the effect of the TEG cold side contacting a worse thermal conductor than itself, an extreme case was tested: a variant of Assembly 1, with aerogel sandwiched between the TEG above and the sand below. It produced what seemed to be a 98° C temperature differential, using the bottom surface of the aerogel as the observed cold side. However, as aerogel acts as a thermal insulator, the same negligible discrepancy that arose when heat moved through conductors in this situation did not occur. Aerogel instead exhibited some kind of thermal bottleneck: the TEG hot side and the lower surface of the aerogel had a 98° C differential, but this dropped to zero when considering the cold side of the TEG directly. Heat pooled within the TEG, as very little energy could trickle through the aerogel insulation; this meant that the two sides of the TEG stabilized at roughly the same temperature. This demonstrates that, to improve the resulting differential, the cold side of the TEG unit must be in contact with a better thermal conductor than itself.

A variant of Assembly 4 where the thermal conductor was solely a modified HTB that ran down to the entire length of the radiator cylinders was also tested, with little significance to the resulting differential. This demonstrates that contact area with soil is more important than sheer solid volume.

While the overall heat capacity of the assembly increases with greater volume, that increase is not significant for high-conductivity materials such as the copper radiator cylinders at these operating temperatures.

This leads into a discussion about optimization of the thermal conductor; any design that optimizes the contact surface area while taking into account other limitations optimizes the resulting temperature differentials. While not tested, conical radiator cylinder assemblies with slopes that match the rate of change of soil conductivity, compared with similarly-designed cylindrical designs that take up the same volume, may potentially maximize the contact area to result in the highest possible differentials.

## V. Limitations And Future Research

Future research could consider more optimized geometries for the radiator cylinders and the HTB, better semiconducting material for the TEG module itself, other methods of applying waste heat from other processes to the TEG module, and an overall more rigorous testing environment. This paper was intended to introduce the conceptual basis behind subsurface assemblies and supplement that idea with simulations and experiments. Future work may expound on the ideas presented in this paper with more strict and exhaustive modelling and experimentation.

Many logistical compromises had to be made throughout the course of this research due to the insufficiency of resources. Namely, compromises had to be made between the cost, quality, and availability of equipment used. Because of time and financial constraints, cheaper and readily-available materials were used. Simulations were run in Fusion 360 which could only support uniform thermal conductivities.

Future work may make use of programs that could model increasing conductivity values and likely achieve higher and more accurate differentials than presented. In addition, simulation/experimental testing could be done using the lower-conductivity lunar soil, which was infeasible for this work due to the constraints of Fusion 360 and the lack of widespread availability for lunar regolith or simulant at this scale.

## VI. Conclusion

A major potential source of energy for surface-level lunar applications comes from thermoelectric technology. Despite their current lower efficiencies in comparison to other available power sources, TEGs are a growing sector in the field of renewable energies, and their use is expected to increase dramatically in the years to come [7, 8]. With the long lifespans of TEGs, such solid-state devices are expected to run long-term without frequent maintenance.

In both an experimental and a simulated environment, the subsurface assemblies were able to create a useful temperature differential across the TEG with a much smaller input heat quantity and environmental temperature differential than exists in actual lunar soil. Through simulations and experimental testing, it was found that increasingly complex assemblies created larger temperature differentials over the plates of TEG units, demonstrating the effectiveness of these devices in harnessing the natural differential of lunar soil. During experimental testing, usable voltages were produced using these temperature differentials, proving that this concept is valid in practical applications [9].

## Acknowledgment

The authors would like to thank Gary Thai for mentorship during the 2021 NASA MINDS program and other members of the authors' original research team who contributed during the program: Marco Lapcevic, Bryan Sanchez, Ryan Phillip, and Alex Gutierrez. The authors would also like to thank: Dr. David Kuijt and Burcu Crothers for administrative assistance; Dr. Led Klosky, Dr. Mustafa Al-Adhami, Dr. Arturo Rankin, Dr. You Zhou, Adam Zeitlin, and various NASA Subject Matter Experts for reviewing the validity and factual accuracy of the research; and Carolyn Siu and Sophia Meytin for technical edits.